\documentstyle[graphicx]{mn}

\def\s{{\rm\thinspace s}}
\def\y{{\rm\thinspace y}}
\def\cm{{\rm\thinspace cm}}
\def\km{{\rm\thinspace km}}
\def\Mpc{{\rm\thinspace Mpc}}
\def\kms{\hbox{$\km\s^{-1}\,$}}

\def\kmpsMpc{\hbox{$\km\s^{-1}\Mpc^{-1}\,$}}

\def\kHz{{\rm\thinspace kHz}}
\def\MHz{{\rm\thinspace MHz}}

\def\refs{\par \noindent \hang}
\def\one_wide{9.0cm}
\def\two_wide{18cm}
\def\hco{HCO$^+$}

\title{New Limits on the Possible Variation of Physical Constants}
\author[M.J. Drinkwater et al.]
{M. J. Drinkwater$^1$,  
J. K. Webb$^1$,
J. D. Barrow$^2$,
V. V. Flambaum$^1$\\
$^1$School of Physics, 
University of New South Wales, Sydney 2052, Australia\\
$^2$Astronomy Centre, University of Sussex, Brighton, BN1 9QH
}
\date{Accepted for publication in MNRAS}

\begin{document}

\maketitle

\begin{abstract}
Recent detections of high-redshift absorption by both atomic hydrogen
and molecular gas in the radio spectra of quasars have provided a
powerful tool to measure possible temporal and spatial variations of
physical `constants' in the Universe.

We have compared the frequency of high-redshift Hydrogen 21\cm\
absorption with that of associated molecular absorption in two quasars
to place new (1 sigma) upper limits on any variation in $y \equiv
g_p\alpha^2$ (where $\alpha$ is the fine structure constant and $g_p$
is the proton g-factor) of $|\Delta y / y| < 5 \times 10^{-6}$ at
redshifts $z=0.25$ and $z=0.68$. These quasars are separated by a
comoving distance of 3000\Mpc\ (for $H_0=75\kmpsMpc$ and $q_0=0$).  We
also derive limits on the time rates of change of $|\dot{g_p}/g_p| < 1
\times10^{-15}\y^{-1}$ and $|\dot{\alpha}/\alpha| <
5\times10^{-16}\y^{-1}$ between the present epoch and $z=0.68$.  These
limits are more than an order of magnitude smaller than previous
results derived from high-redshift measurements.
\end{abstract}


\section{Introduction}
\label{sec_intro}

Quasar absorption systems present ideal laboratories in which to
search for any temporal or spatial variation in the assumed
fundamental constants of Nature. Such ideas date back to the 1930s,
with the first constraints from spectroscopy of QSO absorption systems
arising in the 1960s. An historical summary of the various
propositions is given in Varshalovich \& Potekhin (1995) and further
discussion of their theoretical consequences is given in Barrow \&
Tipler (1986).

The subject of varying constants is of particular current interest
because of the new possibilities opened up by the structure of unified
theories, like string theory and M-theory, which lead us to expect
that additional compact dimensions of space may exist. Although these
theories do not require traditional constants to vary, they allow a
rigorous description of any variations to be provided: one which does
not merely `write in' the variation of constants into formulae derived
under the assumption that they do not vary. This self-consistency is
possible because of the presence of extra dimensions of space in these
theories. The `constants' seen in a three-dimensional subspace of the
theory will vary at the same rate as any change occurring in the extra
compact dimensions. In this way, consistent simultaneous variations of
different constants can be described and searches for varying
constants provide a possible observational handle on the question of
whether extra dimensions exist (Marciano 1984, Barrow 1987, Damour \&
Polyakov 1994).

Prior to the advent of theories of this sort only the time variation of the
gravitational constant could be consistently described using scalar-tensor
gravity theories, of which the Brans-Dicke theory is the simplest example.
The modeling of variations in other `constants' was invariably carried out
by assuming that the time variation of a constant quantity, like the fine
structure constant, could just be written into the usual formulae that hold
when it is constant. The possibility of linked variations in low-energy
constants as a result of high-energy unification schemes has the added
attraction of providing a more powerful means of testing those theories
(Marciano 1984, Kolb, Perry \& Walker 1986, Barrow 1987, Dixit \& Sher
1988, Campbell \& Olive 1995).

Higher-dimensional theories typically give rise to relationships of the
following sort
\begin{eqnarray}
\alpha _i(m_{*}) &=&A_iGm_{*}^2=B_i\lambda ^n(\ell _{pl}/R)^m ;n,m 
{\rm\thinspace constants}  \label{eq.A} \\
\alpha _i^{-1}(\mu ) &=&\alpha _i^{-1}(m_{*}) - \nonumber \\
 \lefteqn{ \pi ^{-1} \sum C_{ij}[\ln(m_{*}/m_j)+\theta (\mu -m_j)\ln (m_j/\mu )]+\Delta _i}  \nonumber 
\end{eqnarray}
where $\alpha _i(..)$ are the three gauge couplings evaluated at the
corresponding mass scale; $\mu $ is an arbitrary reference mass scale,
$m_{*} $ is a characteristic mass scale defining the theory (for
example, the string scale in a heterotic string theory); $\lambda $ is
some dimensionless string coupling; $\ell _{pl}=G^{-1/2}$ is the
Planck length, and $R$ is a characteristic mean radius of the compact
extra-dimensional manifold; $C_{ij} $ are numbers defined by the
particular theories and the constants $A_i$ and $B_i$ depend upon the
topology of the additional ( $>3$) dimensions. The sum is over $j=$
leptons, quarks, gluons, $W^{\pm },\ Z$ and applies at energies above
$\mu \sim 1$\thinspace GeV (Marciano 1984)$.$ The term $\Delta _i$
corresponds to some collection of string threshold corrections that
arise in particular string theories or an over-arching M theory
(Antoniadis \& Quiros 1997). They contain geometrical and topological
factors which are specified by the choice of theory. By
differentiating these two expressions with respect to time (or space),
it is possible to determine the range of self-consistent variations
that are allowed. In general, for a wide range of supersymmetric
unified theories, the time variation of different low-energy constants
will be linked by a relationship of the form (where we consider
$\dot\beta$ to denote the time derivative of $\beta$ etc.)
\begin{equation}
\delta _0\frac{\dot \beta }\beta =\delta _1\frac{\dot G}G+\sum \delta _{2i}
\frac{\dot \alpha _i}{\alpha _i^2}+\delta _3\frac{\dot m_{*}}{m_{*}}+\sum
\delta _{4j}\frac{\dot m_j}{m_j}+\delta _5\frac{\dot \lambda }\lambda +...
\label{eq.B}
\end{equation}
where $\beta \equiv m_e/m_{p}$ is the electron-proton mass ratio. 
It is natural to expect that all the terms
involving time derivatives of `constants' will appear in this relation
unless the constant $\delta $ prefactors vanish because of supersymmetry or
some other special symmetry of the underlying theory. This relation shows
that, since we might expect all terms to be of similar order (although there
may be vanishing constant $\delta $ prefactors in particular theories), we
might expect variations in the Newtonian gravitational `constant', 
$\dot G/G$, to be of order $\dot \alpha /\alpha ^2.$

In this paper, we consider only the bounds that can be placed on the
variation of the fine structure constant and proton $g\ $factor from radio
observations of atomic and molecular transitions in high redshift quasars.
To do this we exploit the recent dramatic increase in quality of
spectroscopic molecular absorption at radio frequencies, of gas clouds at
intermediate redshift, seen against background radio-loud quasars.
Elsewhere, we will consider the implications of simultaneous variations of
several `constants' and show how these observational limits can be used to
constrain a class of inflationary universe theories in which small
fluctuations in the fine-structure constant are predicted to occur.

The rotational transition frequencies of diatomic molecules such as CO
are proportional to ${\hbar /( Ma^2)}$ where $M$ is the reduced mass
and $a={\hbar^2 /( m_e e^2)}$ is the Bohr radius. The
21\textrm{\thinspace cm}\ hyperfine transition in hydrogen has a
frequency proportional to ${\mu_p \mu_B /( \hbar a^3)}$, where $\mu_p
= g_p {e \hbar / (4m_p c)}$, $g_p$ is the proton g-factor and $\mu_B =
{e \hbar / (2m_e c)}$. Consequently 
(the dependence of the results on
the proton-neutron mass difference $m_p-m_n$ is very small so we
assume $m_p/M$ is constant) the 
ratio of a hyperfine frequency to a molecular rotational frequency is
proportional to $g_p \alpha^2$ where $\alpha = {e^2 / ( \hbar c)}$
is the fine structure constant. Any variation in $y\equiv g_p
\alpha^2$ would therefore be observed as a difference in the apparent
redshifts: ${\Delta z / (1+z)} \approx {\Delta y / y}$. Redshifted
molecular emission is hard to detect but absorption can be detected to
quite high redshifts (see review by Combes \& Wiklind 1996). Recent
measurements of molecular absorption in some radio sources
corresponding to known HI 21\cm\ absorption
systems give us the necessary combination to measure this ratio at
different epochs.

Common molecular and HI 21\cm\ absorptions in the radio source PKS
1413+135 have previously been studied by Varshalovich \& Potekhin
(1996). They reported a difference in the redshifts of the CO
molecular and HI 21\cm\ atomic absorptions which they interpreted as a
mass change of $\Delta M / M = (-4 \pm 6) \times10^{-5}$ but as we
show above this comparison actually constrains $g_p \alpha^2$, not
mass.  Furthermore they used overestimates of both the value and
error. They used the Wiklind \& Combes (1994) measurement which had
the CO line offset from the HI velocity by $-$11\kms; a corrected CO
measurement (Combes \& Wiklind 1996) shows there is no measurable
offset.  Furthermore Varshalovich \& Potekhin (1996) used the width of
the HI line for the measurement uncertainty. Even allowing for
systematic errors the true uncertainty is at least a factor of 10
smaller so these data in fact establish a limit of order $10^{-5}$ or
better. This potential for improved limits has prompted the present
investigation: previous upper limits on change in $\alpha$ are of
order $\Delta\alpha/\alpha\approx 10^{-4}$ (Cowie \& Songalia 1995;
Varshalovich, Panchuk \& Ivanchik 1996).

We consider three sources for which combined HI 21\cm\ and molecular
absorption detections have now been published. Before examining them
in detail we look at some examples of local Galactic absorption in
Section~\ref{sec_galactic} to show that the same clouds are detected
in both atomic and molecular absorption. In Section~\ref{sec_high} we
summarize the published measurements of the sources and reanalyze the
HI data with smaller errors than Varshalovich \& Potekhin (1996) to
obtain improved upper limits on the variation of $g_p\alpha ^2$. The
interpretations of the results, a comparison with terrestrial limits,
and future prospects are described in Section~\ref{sec_summary}.

\section{Sources with Galactic absorption}
\label{sec_galactic}

\begin{figure}
\includegraphics[width=\one_wide]{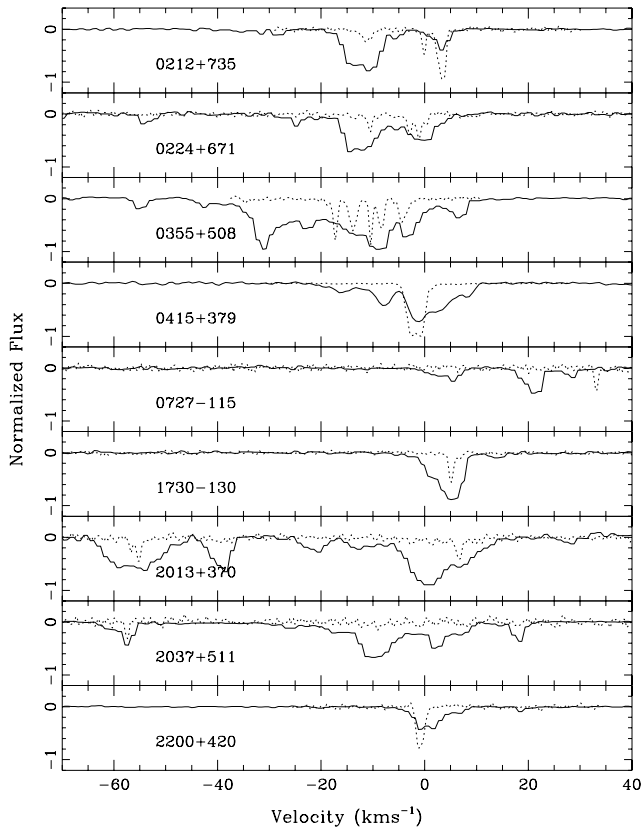}
\centering
\caption{HI and \hco\ spectra of sources with Galactic absorption. The HI
data (solid lines) are from Dickey et al.\ (1983) and the \hco\
data (dotted lines) are from Liszt \& Lucas (1996). The velocity
scale is with respect to the rest frequency of the respective lines
in the local standard of rest frame.}
\label{fig1_spec1}
\label{fig1_spec2}
\end{figure}

\begin{table}
\caption{Sources with Galactic Absorption}
\label{tab_lowz}
\noindent
\begin{tabular}{lrrr}
Name        & NED Name       & l      & b      \\
            &                  & deg    & deg    \\
\\	       					
 0212$+$735 & [HB89] 0212+735  & 128.93 & 11.96  \\ %
 0224$+$671 & 4C +67.05        & 132.12 & 6.23   \\ %
 0355$+$508 & 4C +50.11        & 150.38 & $-$1.60  \\ %
 0415$+$379 & 3C111            & 161.68 & $-$8.82  \\ %
 0727$-$115 & PKS 0727-11      & 227.77 & 3.14   \\ %
 1730$-$130 & [HB89] 1730-130  &  12.03 & 10.81  \\ %
 2013$+$370 & [CCS92] 2013+370 &  74.87 & 1.22   \\ %
 2037$+$511 & [HB89] 2037+511  &  88.81 & 6.04   \\ %
 2200$+$420 & [HB89] 2200+420  &  92.59 & $-$10.44 \\ 
\\
\end{tabular}

Note: `NED Name' gives names under which the sources are recorded 
in the NASA/IPAC Extragalactic Database.
\end{table}

One possible problem with this approach is that the HI and molecular
absorptions might result from different clouds along the respective
lines of sight. We need to show that if we take the nearest HI system
closest in velocity to a given molecular absorption they are
associated and not just a random superposition of different
absorbing clouds.  We investigated this by looking at a sample of
extragalactic mm-wave continuum sources (Liszt \& Lucas 1996). Nine of
these sources have both \hco\ absorption measured by Liszt \& Lucas
and HI absorption measured by Dickey et al.\ (1983).  We note that
these sources are all at low Galactic latitude and will therefore
experience more absorption than we would expect from extragalactic
sources intersecting randomly oriented external galaxies.

The nine Liszt \& Lucas sources with both HI and \hco\ absorption are
listed in Table~\ref{tab_lowz} and we plot their HI spectra with the
\hco\ overlayed in Fig.~\ref{fig1_spec1} using data kindly provided by
the respective authors. They are plotted on common velocity scales
with respect to the rest frequency $f_{rest}$ of the corresponding
transition at the local standard of rest frame. The velocities $v$ are
approximated from the observed frequency $f_{obs}$ by $v/c = 1 -
f_{obs}/f_{rest}$ where $c$ is the speed of light.

The velocity structure of the HI absorption is complex in most of the
sources, consistent with the low Galactic latitudes of these lines of
sight. The tendency for most absorption to be at negative velocities
(as illustrated in Fig.~\ref{fig1_spec1}) is a consequence of Galactic
rotation and the Galactic longitudes of the sources.

Although these lines of sight display many more HI absorption systems
than for \hco, our aim is just to test the hypothesis that every
molecular absorption can be directly associated with HI at the closest
velocity: to do this we fitted velocity components to the HI and
molecular data independently and the compared the velocity differences
statistically.

The procedure adopted for fitting was as follows. As described above,
we treated the HI and \hco\ data independently, so we did not require
the same numbers of components in both HI and \hco\ systems.  We used
the VPFIT software (Webb 1987) to fit multiple components to both the
HI and \hco\ absorption systems.  VPFIT uses an unconstrained
non-linear least-squares optimization method to fit Gaussian/Voigt
profiles to the absorption lines in each system. We adopted the
standard approach of fitting the minimum number of components required
to obtain a statistically acceptable fit, i.e. that the normalized
$\chi^2$ was close to unity. The fitting process was also constrained
not to introduce any components narrower than the instrumental
resolution.

\begin{figure}
\includegraphics[width=\one_wide]{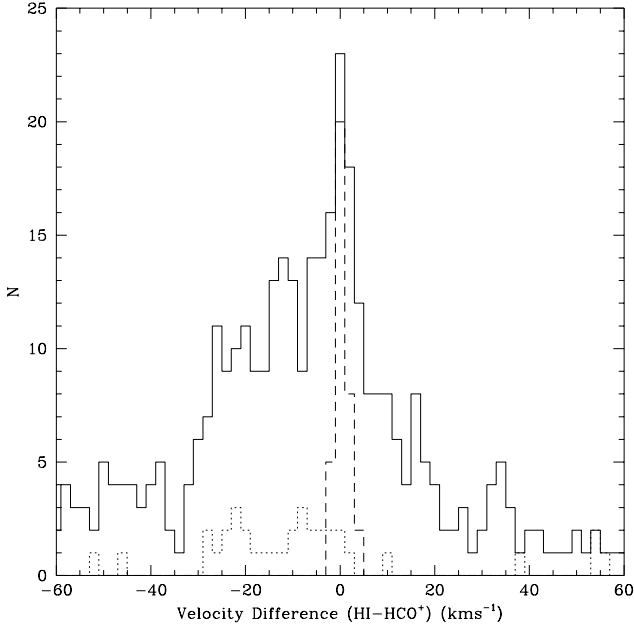}
\caption{Histogram of the HI $-$ \hco\ velocity differences for all
the sources with Galactic absorption listed above. For each source all
line pair differences were used to give the solid histogram. The
dashed line indicates the nearest HI to each \hco\ line. The dotted
line is the result of a random simulation of the nearest pair data to
show that the peak for the nearest lines is real.}
\label{fig2_vhistoa}
\end{figure}

The results of the fitting process are summarized in
Fig.~\ref{fig2_vhistoa}, a histogram of the (HI $-$ \hco) velocity
differences for all systems (solid line) and the closest matching
pairs (dashed line). The distribution of all pairs shows a peak at
zero; it is also slewed to negative velocities because HI is
detected out to larger Galactocentric distances than \hco.

To test our hypothesis of a physical association between each
molecular system and the HI component closest to it in velocity, we
show in Fig.~\ref{fig2_vhistoa} the histogram of closest
(nearest-neighbour) velocity differences. This distribution is centred
close to zero (mean $\Delta v=0.4\kms$) and is very narrow with a
Gaussian dispersion of only 1.2\kms.  This is equal to the spectral
resolution of the HI data (0.6--1.3\kms) (Liszt \& Lucas 1996).  To
show that the narrow distribution of the nearest velocities is
significant and does not result from choosing nearest pairs from
uncorrelated spectra we also plot in Fig.~\ref{fig2_vhistoa} the
distribution of nearest lines (dotted line) where, for each source, we
replaced all the HI velocities by a random selection of HI lines from
all the other sources. This shows no peak at zero so we conclude that
the absorption is from the same clouds to within our velocity
resolution.

\section{High-redshift absorption systems}
\label{sec_high}

\noindent
\begin{table}
\caption{New HI absorption line fits}
\label{fits}
\small
\noindent
\begin{tabular}{cccrl}
Name        &$z_{ref}$   &  $z_{obs}$ &  offset        &  ID  \\
           &             &            &  (\kms)  \\
\\	  
0218$+$357  &.684680$^1$& .684495$\pm$.000010  &$-$32.9$\pm$1.8  &  1\\
     ``    &     ``    & .684568$\pm$.000011  &$-$20.0$\pm$1.9  &  2\\
     ``    &     ``    & .684684$\pm$.000006  &    0.8$\pm$1.0  &  3*\\
1413$+$135 &.246710$^2$& .246710$\pm$.000004  &    0.0$\pm$0.9  &  1*\\
     ``    &     ``    & .246770$\pm$.000037  &   14.5$\pm$8.8  &  2\\
1504$+$377 &.671500$^3$& .673015$\pm$.000014  &  271.6$\pm$2.5  &  1\\
     ``    &     ``    & .673236$\pm$.000003  &  311.2$\pm$0.5  &  2*\\
     ``    &     ``    & .673423$\pm$.000005  &  344.7$\pm$0.8  &  3*\\
     ``    &     ``    & .673569$\pm$.000068  &  370.8$\pm$12   &  4\\
\\
\end{tabular}

Notes: $z_{obs}$ is the fitted redshift and uncertainty including both fitting
and velocity scale (0.3\kms) errors; offset is the velocity of the
fitted line with respect to the reference redshift $z_{ref}$ taken from
the references listed in Table~\ref{published}. ID is 
the number of the component labeled in Fig.~\ref{fig5_spec} with an *
to indicate the strongest lines used to compare with the molecular data.
\end{table}

\subsection{Sample of redshifted absorbers}
\label{sec_sample}

We found three sources with common extragalactic molecular and HI
absorption detections published: 0218$+$357, 1413$+$135 and
1504$+$377. The details of these sources are listed below in
Table~\ref{published}. The sources include 1413$+$135 which was
previously discussed by Varshalovich \& Potekhin (1996).  The
published HI data are also included in Table~\ref{published}. The
typical HI error of $\Delta z / (1+z) = 1\times10^{-5}$ (3\kms) is
smaller than the value used by Varshalovich \& Potekhin (1996), but
still the dominant term. We show in the next section how we improved
the HI precision by refitting HI lines to the data kindly provided by
Dr. C. Carilli. We then compared these to the published molecular data
to put new limits on the variation of $g_p\alpha ^2$.

\subsection{HI absorption systems}
\label{sec_hi}

\begin{figure}
\includegraphics[width=\one_wide]{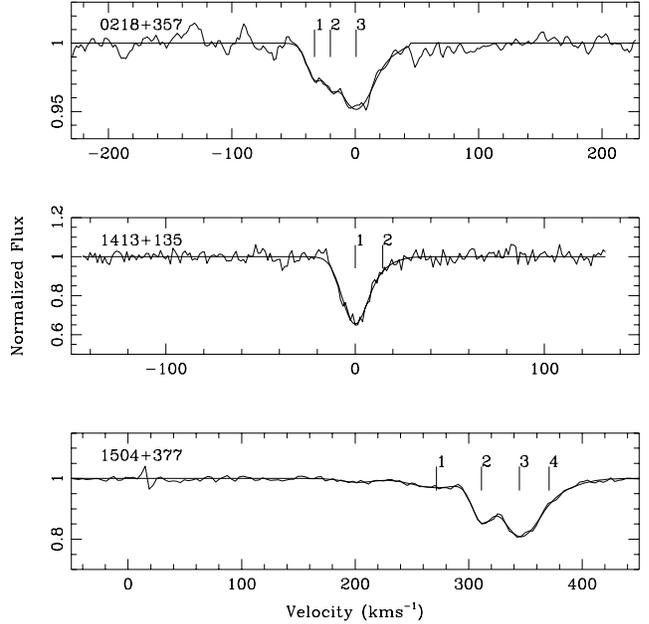}
\caption{Spectra of three quasars with high-redshift HI absorption.
The data are from references 2, 4 and 6 listed in
Table~\ref{published}. The velocity zero points are taken from the
same references.
The spectra were first normalized to a
continuum of unity and then deconvolved by fitting multiple absorption
components as described in the text: the final fit is plotted over the
data as a continuous smooth line with the centres of each component
marked.}
\label{fig6_fits}
\label{fig5_spec}
\end{figure}

In this section we describe how we obtained more precise redshifts for
the HI absorption data. A good rule of thumb is that the uncertainty
in fitting a line position is about one tenth of a resolution
element. In the case of the HI data for 1504$+$377 the resolution was
10 \kHz\ per channel at 848.8 \MHz\ or about 3\kms, about the error
quoted by Carilli et al.\ (1997). Applying the one tenth of a channel
criterion we might expect to better than this, although we must also
consider any systematic errors. Checks made by Carilli et al.\ (1997)
indicate that the systematic errors in the velocity scale are about
0.8 \kHz\ or 0.3\kms.

The HI data for three sources are shown plotted in
Fig.~\ref{fig5_spec}.  We used to the VPFIT software described above
to decompose the absorption systems, following the same procedure as
before, fitting a number of components to each system. The resulting
fits are plotted over the data in Fig.~\ref{fig5_spec}. All the
components we fitted to each source are listed in Table~\ref{fits},
both in terms of observed redshift and a velocity offset with respect
to the nominal redshift.  The fitting routine gave uncertainties less
than 1\kms\ for the strong lines; these were added in quadrature with
the velocity scale uncertainty of 0.3\kms\ to obtain the errors given
in Table~\ref{fits}.

As with the low redshift galactic absorption we took the nearest
matches between these HI absorption detections and the
molecular absorption redshifts listed in Table~\ref{published}.

\begin{table*} 
\centering
\begin{minipage}{160mm}
\caption{Comparison of Molecular and Atomic Absorption Data}
\label{published}
\noindent
\begin{tabular}{lcrrrrr}
Name      & NED Name&Molecular z            & Molecule & Atomic z (HI)   &Atomic z (HI, new)& $\sigma_{\Delta z / 1+z}$ \\
\\
0218$+$357& TXS 0218+357&$0.684680\pm0.000006^1$&\hco (2-1)& 0.68466$\pm0.00004^2$ &$0.684684\pm0.000006$ &$5\times10^{-6}$ \\
1413$+$135& PKS 1413+135&$0.246710\pm0.000005^3$& CO(0-1)  & 0.24671$\pm0.00001^4$ &$0.246710\pm0.000004$ &$5\times10^{-6}$ \\
1504$+$377& B3 1504+377 &$0.673350\pm0.000005^5$&\hco (2-1)& 0.67324$\pm0.00001^6$ &$0.673236\pm0.000003$& --\\
1504$+$377& ``&$0.673350\pm0.000005^5$&\hco (2-1)& 0.67343$\pm0.00001^6$ &$0.673423\pm0.000005$& --\\
\\
\end{tabular}
References:\\
1: Wiklind \& Combes (1995) modified; see text,\\
2: Carilli, Rupen \& Yanny (1993),\\
3: Wiklind \& Combes (1994) corrected by Combes \& Wiklind (1996),\\
4: Carilli, Perlman \& Stocke (1992),\\
5: Wiklind \& Combes (1996),\\
6: Carilli et al.\ (1997).
\end{minipage}
\end{table*}

\subsection{Molecular absorption systems}
\label{sec_mol}

The redshifts of the molecular absorptions are listed in
Table~\ref{published}. These are---in principle---much more precise
than the HI data, with velocity uncertainties of 0.1\kms.  In
practice, the published values were not quoted to this precision
because they were not aimed at this kind of work. We have taken the
best estimates of the molecular redshifts and uncertainties from the
publications as follows.

In the case of 0218$+$357 Wiklind \& Combes (1995) work with respect
to the HI redshift of $z=0.68466$ (Carilli, Rupen \& Yanny 1993) but
it is clear from their Fig.~3 that the centre of the molecular
absorption (measured from the \hco\ system) is $4\pm0.5$\kms\ higher
corresponding to $z=0.684680\pm0.000006$, the value we adopt here. 

The velocity of the CO absorption in 1413$+$135 was originally found
to be offset from the HI result by Wiklind \& Combes (1994) but this
was subsequently corrected and is equal to the HI value of z=0.24671
(Combes \& Wiklind 1996). We assume the uncertainty in the molecular
redshift is half the least significant digit, although the
potential value is much smaller.

The redshift of the molecular absorption in 1504$+$377 quoted by
Wiklind \& Combes (1996) (z=0.67335) is not consistent with the
HI detections of (Carilli et al.\ 1997).  Carilli et al.\ interpret
this difference as either a shift of 15\kms\ in one of the velocity
scales or intrinsic differences in the atomic and molecular
absorption. Their HI redshift has been confirmed by subsequent
independent observations (Carilli et al.\ 1998, Ap.J., submitted). 
Our own deconvolution of the HI data gives a 4-line
system very similar to, but displaced by a constant shift from, the
\hco\ absorption data in Fig.~2(a) of Wiklind \& Combes (1996). An
offset of a single line could be interpreted as a case of different
clouds causing the absorption, but a systematic offset of 4 lines is
much more likely due to a velocity error in one of the measurements.
We have not used 1504$+$377 in our analysis below because of this
disagreement.

\subsection{Comparison of HI and molecular systems}
\label{sec_analyse}

Our new more accurate redshift estimates are listed with the molecular
results in Table~\ref{published}. In the two sources 0218$+$357 and
1413$+$135 the redshifts agree to within our errors. (As discussed
above we are not analyzing 1504$+$377.)  We can therefore combine the
uncertainties in quadrature to give 1-sigma upper limits on the
redshift differences.  These are shown in the Table and are
$|{\Delta z / 1+z}| < 5\times10^{-6}$ (1.5\kms) for both sources.

We must still consider the possibility that the molecular and atomic
absorption arises in different gas clouds along the line of sight.
This could explain any observed difference. However there is no
measurable difference between the two velocities in our data, so we
are probably detecting the same gas. The alternative would be that
there was a change in the frequencies but that in both cases it was
exactly balanced by the random relative velocity of the two gas clouds
observed. We consider this very unlikely because of the small 1\kms\
dispersion within single clouds shown in Section~\ref{sec_galactic}.

\section{Discussion}
\label{sec_summary}

We can now use the limits to ${\Delta z / 1+z}$ with the relationship
given in Section~\ref{sec_intro} to derive 1-sigma limits on any
change in $y=g_p \alpha^2$: $|{\Delta y / y}| < 5\times10^{-6}$ at
both $z=0.25$ and $z=0.68$. These are significantly lower than the
previous best limit of $1\times10^{-4}$ by Varshalovich \& Potekhin
(1996) (it was quoted as a limit on nucleon mass, but it actually
refers to $g_p \alpha^2$).

As there are no theoretical grounds to expect that the changes in
$g_p$ and $\alpha^2$ are inversely proportional, we obtain independent
rate-of-change limits of $|\dot{g_p}/g_p| < 2\times10^{-15}\y^{-1}$
and $|\dot{\alpha}/\alpha| < 1\times10^{-15}\y^{-1}$ at
$z=0.25$ and $|\dot{g_p}/g_p| < 1 \times10^{-15}\y^{-1}$ and
$|\dot{\alpha}/\alpha| < 5\times10^{-16}\y^{-1}$ at $z=0.68$
(for $H_0=75\kmpsMpc$ as adopted throughout this section).  These new
limits are much lower than the previous 1 sigma limit of
$|\dot{\alpha}/\alpha| < 8\times10^{-15}\y^{-1}$ at $z\approx
3$ (Varshalovich et al.\ 1996).

The most stringent laboratory bound on the time variation of $\alpha $
comes from a comparison of hyperfine transitions in Hydrogen and
Mercury atoms (Prestage, Tjoelker \& Maleki 1995), $\left| \dot \alpha
/\alpha \right| <3.7\times 10^{-14}\y^{-1},$ and is significantly
weaker than our astronomical limit.  The other strong terrestrial
limit that we have on time variation in $\alpha $ comes from the
analysis of the Oklo natural reactor at the present site of an
open-pit Uranium mine in Gabon, West Africa. A distinctive thermal
neutron capture resonance must have been in place 1.8 billion years
ago when a combination of fortuitous geological conditions enriched
the subterranean Uranium-235 and water concentrations to levels that
enabled spontaneous nuclear chain reactions to occur (Maurette
1972). Shlyakhter (1976, 1983) used this evidence to conclude that the
neutron resonance could not have shifted from its present
specification by more than $5\times 10^{-4}${\thinspace eV} over the
last 1.8 billion years and, assuming a simple model for the dependence
of this energy level on coupling constants like $\alpha ,$ derived a
limit in the range of $\left| \dot \alpha /\alpha \right| <
(0.5$--$1.0)\times 10^{-17}\y^{-1}.$ The chain of reasoning leading to
this very strong bound is long, and involves many assumptions about
the local conditions at the time when the natural reactor ran,
together with modeling of the effects of any variations in
electromagnetic, weak, and strong couplings. Recently, Damour \& Dyson
(1997) have provided a detailed reanalysis in order to place this
limit on a more secure foundation. They weaken Shlyakhter's limits
slightly but give a 95\% confidence limit of $-6.7\times
10^{-17}\y^{-1}<\dot \alpha /\alpha <5.0\times 10^{-17}\y^{-1}.$
However, if there exist simultaneous variations in the electron-proton
mass ratio this limit can be weakened.

The Oklo results provide stronger limits on the time variation of
$\alpha $ than the astronomical data; however, the astronomical limits
have the distinct advantage of resting upon a very short chain of
theoretical deduction and are more closely linked to repeatable
precision measurements of a simple environment. The Oklo environment
is sufficiently complex for significant uncertainties to remain.

Unlike the Oklo limits, the astronomical limits also allow us to
derive upper limits on any {\em spatial} variation in $\alpha .$
Spatial variation is expected from the theoretical result that the
values of the constants would depend on local conditions and that they
would therefore vary in both time and space (Damour \& Polyakov 1994).
The two sources for which we derived limits, 0218$+$357 and
1413$+$135, are separated by 131 degrees on the sky, so together with
the terrestrial result, we find the same values of $\alpha$ to within
$|{\Delta \alpha / \alpha}| < 3\times10^{-6}$ in three distinct
regions of the universe separated by comoving separations up to
3000\Mpc. Limits on spatial variation of $g_p \alpha^2 m_e/m_p$ were
previously discussed by by Pagel (1977, 1983) and Tubbs \& Wolfe
(1980). We have improved on their limits by some 2 orders of magnitude
but as our sources are at lower redshift, they are not causally
disjoint from each other.

With a careful reanalysis of the available data we have made a
significant improvement on the existing upper limits on any change of
the product of two physical constants $g_p\alpha ^2$ with cosmic
time. The high-redshift measurements are now approaching the best
terrestrial measurements based on the Oklo data. These could be
further improved by a factor of 2--5 with additional observations that
would not be difficult to perform such as fitting the atomic and
molecular data simultaneously, remeasuring the HI absorptions at
higher spectral resolution, and checking the discrepancy in the
observations of 1504$+$377.

\section*{Acknowledgments}

We are very grateful Drs. C. Carilli, H. Liszt, and J. Dickey for many
helpful discussions and for allowing us to use their data. We also
wish to thank Drs. F. Combes, T. Wiklind and A. Potekhin for helpful
comments about this work. The referee of this paper also made a number
of suggestions that greatly improved the presentation of these
results.  We used the NASA/IPAC Extragalactic Database (NED) which is
operated by the Jet Propulsion Laboratory, Caltech, under contract
with the National Aeronautics and Space Administration.

\section*{References}

\refs Antoniadis, I., Quiros, M., 1997, Phys. Lett., B392, 61
\refs Barrow, J. D., 1987, Phys. Rev., D35, 1805
\refs Barrow, J. D., Tipler, F. J., 1986, The Anthropic Cosmological Principle. Oxford Univ. Press, Oxford, p. 533
\refs Campbell, B. A., Olive, K. A., 1995, Phys. Lett., B345, 429
\refs Carilli, C. L., Perlman, E. S., Stocke, J. T., 1992, ApJ, 400, L13
\refs Carilli, C. L., Rupen, M. P., Yanny, B., 1993, ApJ, 412, L59
\refs Carilli, C. L., Menten, K. M., Reid, M. J., Rupen, M. P., 1997, ApJ, 474, L89
\refs Combes, F., Wiklind, T., 1996, in Bremer, M., van der Werf, P., Rottgering, H., Carilli, C., eds, Cold Gas at High Redshift. Kluwer, Dordrecht, p. 215
\refs Cowie, L. L., Songalia, A., 1995, ApJ, 453, 596
\refs Damour, T., Dyson, F., 1997, Nucl. Phys., B480, 37
\refs Damour, T., Polyakov, A. M., 1994, Nucl. Phys., B423, 596
\refs Dickey, J. E., Kulkarni, S. R., van Gorkom, J. H., Heiles, C. E., 1983, ApJS, 53, 591
\refs Dixit, V. V., Sher, M., 1988, Phys. Rev., D37, 1097
\refs Kolb, E. W., Perry, M. J., Walker, T. P., 1986, Phys. Rev., D33, 869
\refs Liszt, H., Lucas, R., 1996, A\&A, 314, 917
\refs Marciano, W., 1984 Phys. Rev. Lett., 52, 489
\refs Maurette, M., 1972, Ann. Rev. Nucl. Part. Sci., 26, 319
\refs Pagel, B. E. J., 1977, MNRAS 179, 81P
\refs Pagel, B. E. J., 1983, Phil. Trans. Roy. Soc., A310, 245
\refs Prestage, J. D., Tjoelker, R. L., Maleki, L., 1995, Phys. Rev. Lett., 74, 3511
\refs Shlyakhter, A. I., 1976, Nature, 264, 340
\refs Shlyakhter, A. I., 1983, Direct test of the time-independence of fundamental nuclear constants using the Oklo natural reactor, ATOMKI Report A/1
\refs Tubbs, A. D., Wolfe, A. M., 1980, ApJ, 236, L105
\refs Varshalovich, D. A., Potekhin, A. Y., 1995, Space Science Review, 74, 259
\refs Varshalovich, D. A., Potekhin, A. Y., 1996, Pis'ma Astron. Zh., 22, 3
\refs Varshalovich, D. A., Panchuk, V. E., Ivanchik, A. V., 1996, Astron. Lett., 22, 6
\refs Webb, J. K., 1987, PhD thesis, University of Cambridge
\refs Wiklind, T., Combes, F., 1994, A\&A, 286, L9
\refs Wiklind, T., Combes, F., 1995, A\&A, 299, 382
\refs Wiklind, T., Combes, F., 1996, A\&A, 315, 86

\end{document}